\documentclass[conference,letterpaper,final,10pt]{IEEEtran}
\usepackage[noadjust]{cite}
\usepackage{amsmath}
\usepackage[pdftex]{graphicx}
\usepackage{caption}
\usepackage{subcaption}

\usepackage{amsfonts}
\usepackage{amssymb}

\newcommand{\figref}[1]{Fig.~\ref{fig:#1}}
\newcommand{\tabref}[1]{Table~\ref{tab:#1}}
\newcommand{\chapref}[1]{\ref{sec:#1}}

\newcommand{\Exp}[2]{\mathbb{E}_{#1}\!\left\{#2\right\}}
\newcommand{\Pnull}{\textbf{P}^{\perp}_{\textbf{H}}}
\newcommand{\C}{\mathbb{C}}
\newcommand{\N}{\mathcal{N}}
\newcommand{\NC}{\mathbb{C}\mathcal{N}}
\newcommand{\ydl}{\textbf{r}_{\text{DL}}}

\newcommand{\xdl}{\textbf{x}_{\text{DL}}}

\newcommand{\Gdl}{\textbf{G}_{\text{DL}}}

\newcommand{\EVMDL}{\text{EVM}_{\text{DL}}}
\newcommand{\Edl}{\textbf{E}_{\text{tx}}}

\IEEEoverridecommandlockouts
\begin{document}

\title{\LARGE On the Impact of Hardware Impairments on Massive MIMO}

\author{\IEEEauthorblockN{Ulf Gustavsson\IEEEauthorrefmark{1}, Cesar Sanch\'ez-Perez\IEEEauthorrefmark{4}, Thomas Eriksson\IEEEauthorrefmark{3}, Fredrik Athley\IEEEauthorrefmark{2}, \\ Giuseppe Durisi\IEEEauthorrefmark{3}, Per Landin\IEEEauthorrefmark{3}, Katharina Hausmair\IEEEauthorrefmark{3}, Christian Fager\IEEEauthorrefmark{4} and Lars Svensson\IEEEauthorrefmark{5}
}
\IEEEauthorblockA{\IEEEauthorrefmark{1}Ericsson AB}
\IEEEauthorblockA{\IEEEauthorrefmark{2}Ericsson Research}
\IEEEauthorblockA{\IEEEauthorrefmark{3}Communication Systems Group, Dept. of Signals and Systems, Chalmers University of Technology}
\IEEEauthorblockA{\IEEEauthorrefmark{4}Microwave Electronics Laboratory, Dept. of Microtechnology and Nanoscience, Chalmers University of Technology}
\IEEEauthorblockA{\IEEEauthorrefmark{5}Computer Engineering Division, Department of Computer Science and Engineering, Chalmers University of Technology}
\thanks{This work was partly supported by the Swedish Foundation for Strategic Research under grant SM13-0028.}
}
\maketitle

\begin{abstract}
Massive multi-user (MU) multiple-input multiple-output (MIMO) systems are one possible key technology for next generation wireless communication systems. Claims have been made that massive MU-MIMO will increase both the radiated energy efficiency as well as the sum-rate capacity by orders of magnitude, because of the high transmit directivity. However, due to the very large number of transceivers needed at each base-station (BS), a successful implementation of massive MU-MIMO will be contingent on of the availability of very cheap, compact and power-efficient radio and digital-processing hardware. This may in turn impair the quality of the modulated radio frequency (RF) signal due to an increased amount of power-amplifier distortion, phase-noise, and quantization noise.

In this paper, we examine the effects of hardware impairments on a massive MU-MIMO single-cell system by means of theory and simulation. 
The simulations are performed using simplified, well-established statistical hardware impairment models as well as more sophisticated and realistic models based upon measurements and electromagnetic antenna array simulations.
\end{abstract}

\IEEEoverridecommandlockouts
\IEEEpeerreviewmaketitle


\section{Introduction}

The potential gains of massive MU-MIMO have so far been assessed mainly through theoretical analysis \cite{LarssonMaMIMONextGen,RusekScaling}. It has been claimed that massive MU-MIMO may increase the radiated power efficiency in the range of thousands, while multiplying the sum-rate capacity tens of times. Both claims rely on massive MIMO unprecedented capability of focusing the transmitted power  in space, which enables spatial user separation and makes it possible to serve all users within the same time-frequency resource. 
Due to the very large array dimensions, massive MU-MIMO relies heavily on the availability of cheap and power efficient radio and base-band hardware. 
However, decreasing the hardware cost and increasing its power efficiency may yield significant side effects, such as distortions and noise. These impairments may in turn have a negative impact on the overall system performance.

Hardware impairments and their impact on massive MU-MIMO have been examined using simplified stochastic models in e.g., \cite{StuderHWImpairments,BjornsonHW3,BjornsonHW1,ZhangMBCD14}. These models commonly rely on the assumption that the transmit waveform and the hardware impairments are uncorrelated. But since most impairments such as power amplifier distortion, phase noise, and quantization noise, are highly dependent on the transmit waveform, more refined analyses are called for. 
It is worth noting that standard algorithms for hardware-impairment compensation, such as digital pre-distortion (DPD) \cite{Bassam_Helaoui_Ghannouchi09,MIMODPDCombFB} or phase-noise estimation and compensation \cite{bittner2009oscillator} may be too complex in a massive MU-MIMO system, due to the very large array size.

Another source of impairments is the mutual coupling between antenna ports, which changes the nominal load-impedance of the power amplifiers and cause additional distortion. Some methods have been proposed for the compensation of mutual coupling-induced distortion in e.g., \cite{Bassam_Helaoui_Ghannouchi09,MIMODPDCombFB}, but these methods are too complex for massive MU-MIMO.

Although in a massive MU-MIMO system the output power per antenna decreases as one over the number of antennas, the power consumed by the base-band hardware and by the data converters grows linearly with the array size. 
This may be mitigated by decreasing the digital resolution, i.e., the number of bits used by the data converters, which yields a reduction in power consumption according to Walden's figure of merit~\cite{walden1999analog},\footnote{Walden's figure is defined for analogue-to-digital converters (ADC), but applies in the same manner to DAC's.} but also increases the quantization noise.

In this paper, we  examine the impact of two key impairments, i.e., power amplifier (PA) distortion, and digital-to-analogue converter (DAC) quantization noise, on the performance of a massive MU-MIMO base-station (BS) in a single-cell scenario with multiple users, under different channel conditions.

\paragraph*{Notation}
Throughout the paper, $M$ denotes the number of transmit-antenna ports and $K$ denotes the number of users. Vectors and matrices are in bold typeface, e.g., $\textbf{x}$ and $\textbf{X}$, respectively, while scalars are in regular typeface, e.g., $x$.
With $\|\textbf{X}\|_F = \sqrt{\text{Tr}\left(\textbf{X}^H\textbf{X}\right)}$ we indicate the Frobenius norm of $\textbf{X}$.
Here, $\textbf{X}^H$ stands for the conjugate transpose of the matrix $\textbf{X}$ and $\text{Tr}\left(\cdot\right)$ denotes the trace operator. 
For complex-valued $\textbf{x}$ or $x$, we  denote their conjugate as $\bar{\textbf{x}}$ and $\bar{x}$ respectively.  
The normal distribution with mean $\mu$ and variance $\sigma^2$ is denoted by $\N\left(\mu,\sigma^2\right)$, while $\NC\left(\textbf{0},\textbf{C}\right)$ stands for the distribution of a circularly symmetric complex normal random variable with covariance matrix~$\textbf{C}$. 
Throughout the paper, we study the MU-MIMO down-link (DL) channel described by
\begin{equation}
	\ydl = \textbf{H}^H\xdl + \textbf{n}.
\end{equation}
Here, $\ydl\in\C^K$ is the received signal vector; $\textbf{H}\in\C^{M\times K}$ is the channel matrix; $\textbf{n}\in\C^K$ is the receive noise; $\xdl\in\C^M$ is the pre-coded data vector resulting from $\xdl = \Gdl\textbf{s}$ where $\textbf{s}\in\C^K$ is the vector of transmit data symbols and $\Gdl\in\C^{M\times K}$ is the pre-coding matrix, which relies on the channel knowledge available at the BS.

\paragraph*{Paper Outline}
The paper is structured as follows. In Section \chapref{simulator} we  introduce and discuss both stochastic and knowledge-based, deterministic models for some of the key hardware impairments. A discussion on performance metrics then follows in Section~\chapref{metrics}, after which a set of parametric studies are presented and discussed in Sections~\chapref{Paramstudy} and~\chapref{Combined}. A summary and some concluding remarks are given in Section~\chapref{conclusions}.


\section{Hardware Impairment Models}
\label{sec:simulator}

Radio and digital hardware are imperfect contraptions that inflict different types of distortions onto the desired transmit signal. In this section, we review two simple stochastic approaches to model hardware imperfections. We then present a more accurate deterministic model, which is used for comparative simulations in Section~\chapref{Paramstudy}.

\subsection{Additive Stochastic Impairment Models}
\label{sec:addmod}

One commonly adopted model for hardware (HW) impairments in the massive MU-MIMO literature is the additive stochastic impairment model, in which the residual impairments after compensation are treated as additive Gaussian noise. A modified version of this additive model, described in its original form in, e.g., \cite{BjornsonHW3}, yields the following DL input-output relation
\begin{equation}
	\ydl = \alpha\textbf{H}^H(\xdl + \textbf{w}) + \textbf{n}
	\label{eq:AddMod}
\end{equation}
where the modification consists of the normalization constant
\begin{equation}
	\alpha = \sqrt{\frac{\|\xdl\|^2_F}{\|\xdl + \textbf{w}\|^2_F}}.
	\label{eq:AddModNorm}
\end{equation}
This constant ensures that no energy is added by the nonlinearities of the system, as dictated by the Manley-Rowe relation \cite{ManleyRowe}. The impairment noise is defined as $\textbf{w}\sim\NC\left(\textbf{0},\textbf{C}_w\right) $, where $\textbf{C}_w = \nu\cdot\text{diag}\left(W_{1,1},\ldots,W_{M,M}\right)$, with $W_{m,m}$ being the $M$ diagonal elements of the covariance matrix $\textbf{C}_{\xdl} = \Exp{}{\xdl\xdl^H}$ and where $\nu$ is a proportionality constant~\cite{BjornsonHW3}. 

This model, whose main feature is that the additive impairment depends on the transmit signal only through its covariance matrix, is appealing because it is analytically tractable. However, its accuracy is questionable, because most hardware impairments are amplitude dependent. This leads us to the multiplicative stochastic models.

\subsection{Multiplicative Stochastic Impairment Models}
\label{sec:mulmod}
A different approach is to model the stochastic impairments as multiplicative amplitude and phase errors, \cite{AthleyHW}. This yields an amplitude-dependent error term, and results in the following DL input-output relation
\begin{equation}
	\ydl =  \beta\textbf{H}^H \Edl \xdl + \textbf{n}.
	\label{eq:MultMod}
\end{equation}
where $\Edl = \text{diag}\left(\{(1 + a_m)\exp\left(-i\phi_m\right)\}_{m=1}^M \right)\in\C^{M\times M}$ is the multiplicative error matrix, with $a_m\sim\mathcal{N}(0,\sigma^2_a)$ and $\phi_m\sim\mathcal{N}(0,\sigma^2_{\phi})$ being the stochastic amplitude and phase errors respectively. Here,
\begin{equation}
	\beta = \sqrt{\frac{\|\textbf{H}^H\|^2_F}{\|\textbf{H}^H \Edl\|^2_F}}
	\label{eq:MultModNorm}
\end{equation}
is a normalization constant applied to ensure that energy conservation holds.

From a HW impairment point of view, this modeling approach appears more sound than its additive counterpart, since both PA distortion and phase noise are by nature amplitude dependent. The quantization noise from data-converters is, however, additive and strongly dependent on the signal.

Both modeling approaches are simplistic and may not fully capture the impact of hardware impairments. This motivates us to introduce more refined deterministic behavioral models.

\subsection{Deterministic Behavioral Models}

We provide next accurate models for some of the main sources of hardware impairments in massive MU-MIMO systems, i.e., the power amplifier, the antenna array, and the data converters.  These models will allow us i) to characterize the impact of hardware impairments through a simulation study presented in Section~\chapref{Paramstudy}, and ii) to assess the accuracy of the stochastic  models reviewed in Sections \chapref{addmod} and \chapref{mulmod}.

\subsubsection{The Power Amplifier and Antenna Array Models}
\label{sec:PAMod}

Power amplifiers (PAs) are commonly modeled by Volterra-series~\cite{Schetzen80}, or subsets of these, e.g., memory polynomials~\cite{Morgan_Ma_Kim_Zierdt_Pastalan06}. These models describe accurately radio frequency (RF) PAs in a 50 $\Omega$ environment. However, they fail to capture mutual coupling and mismatch effects.

The effect of mismatch and mutual coupling from the neighboring PAs can be modeled as described in \cite{FagerPA}, where the output of the $m$th power amplifier is expressed as
\begin{equation}
	y_{m}[n] = f(x_{m}[n],x_{m,r}[n])
	\label{eq:PAMod}
\end{equation}
where $x_{m}[n]$ is the input signal and $x_{m,r}[n]$ the sum-coupled signal at the $m$th PA output stemming from the other $M-1$ antennas through the linear relation defined by the scattering parameter (S-parameter) matrix. Here, $f(\cdot)$ is a non-linear memory-polynomial function involving different non-linear combinations of $x_{m}[n]$  and $x_{m,r}[n]$

This dual-input modeling strategy has been verified through wideband load-pull measurements \cite{FagerPA} and will be used in the simulations presented in this paper. The total far-field component of the array is calculated by superposition
\begin{equation}
	\bar{E}_{tot}(\theta,\varphi)[n] = \sum_{m = 1}^{M} y_{m}[n] E_m(\theta,\varphi)
	\label{eq:Farfield}
\end{equation}
where $M$ is the number of antennas and $E_m(\theta,\varphi)$ is the distance-normalized far-field pattern from the $m$th antenna element when this element is excited and all other elements are terminated in 50~$\Omega$. The scalars $\theta$ and $\varphi$ are the elevation and azimuth angles, respectively. 

The antenna model is simulation-based with S-parameters generated using the 3D EM simulator Ansys HFSS, \cite{AnsysHFSS}. This provides the degree of coupling between elements, which is used to compute the reflection component $x_{m,r}[n]$ in \eqref{eq:PAMod}, as described in~\cite{FagerPA}. With this approach, different antenna array configurations with different number of elements and spacing can be easily and accurately analyzed.

\subsubsection{Data Converter Models}
\label{sec:DCMods}

In this paper, we consider data-converters that use uniform quantization with ideal timing, i.e. no jitter or timing errors. Uniform quantization is performed using a pre-defined number of bits over a pre-determined dynamic range above which the converter performs hard-limiting. For the complex case, we use a Cartesian quantization scheme that performs quantization independently for real and  imaginary parts.

%


\section{Simulation Setup and Performance Metrics}
\label{sec:metrics}

We consider single carrier quadrature amplitude modulation (QAM) with spatial multiplexing, e.g., all users use the same spectrum. The simulations are performed in complex base-band and, in order to analyze the nonlinear PA distortion, the signal is oversampled. The simulation parameters used throughout this paper (unless otherwise stated) are summarized in \tabref{SimParams}.
\begin{table}[t]
\centering
		  \caption{Simulation parameters.}
  \label{tab:SimParams}
	\begin{tabular}{|c|c|}
    \hline
    \textbf{Parameter} & \textbf{Value} \\ \hline\hline
    QAM order & 64 \\ \hline
    Pulse-shaping filter & Root-Raised Cosine \\ \hline
		Roll-off factor & 0.22 \\ \hline
		Oversampling Ratio (OSR) & $\times 5$ \\ \hline
    SNR & 10 dB \\ \hline
    Number of antennas $(M)$ & 4 to 225 \\ \hline
		Array configuration & Rectangular \\ \hline
		Antenna type & Rectangular patch\\ \hline
		Number of users $(K)$ & 4 \\ \hline
		UE Distribution & Uniform  over\\ 
		 & $[-60,60]$ deg. azimuth and\\
		 & $[-30,30]$ deg. elevation \\ \hline
    \end{tabular}
\end{table}

In order to assess the performance degradation resulting from different types of hardware impairments, one needs easily quantifiable metrics. 
In this paper, we choose as metrics the error vector magnitude (EVM) averaged over the users as well as the unwanted space-frequency emissions.

\subsection{Error Vector Magnitude}

The EVM in the DL is defined as
\begin{equation}
	\EVMDL = 100\cdot\sqrt{\frac{\Exp{}{\|\textbf{s}-\rho\ydl\|_2^2}}{\Exp{}{\|\textbf{s}\|_2^2}}}
	\label{eq:EVMDef}
\end{equation}
where $\rho = \ydl^H\textbf{s} / \|\ydl\|_2^2$ is a scaling factor that removes constant gain and phase errors, and $\textbf{s}\in\C^K$ is the DL data-vector before pre-coding. The EVM is a comprehensive figure of metric because it takes noise, distortion, and interference into account.

\subsection{A Metric for Unwanted Spatial Emissions}

Hardware impairments may cause unwanted emissions on adjacent channels. For SISO transmitters, these unwanted emissions are typically characterized in terms of adjacent channel leakage ratio (ACLR) measured at the antenna reference port. ACLR is the ratio between the desired in-band power and the out of band unwanted emission power. This figure is, however, only defined over frequency and not over space, hence providing only a partial characterization of unwanted emissions in multiple antenna systems. Our approach is to integrate the transmitted power over both frequency and space. The adjacent space-frequency leakage ratio (ASLR) is computed by integrating the total far-field component over space and frequency around the intended user as
\begin{equation}
	\text{ASLR} = \frac{\displaystyle\iiint\limits_{\Omega_{\text{Useful}}} \left|\sum_{m = 1}^{M} Y_{m}(\omega) E_m(\theta,\varphi) \right|^2 \ d\omega d\theta d\varphi}
	{\displaystyle\iiint\limits_{\Omega_{\text{Unwanted}}} \left|\sum_{m = 1}^{M} Y_{m}(\omega) E_m(\theta,\varphi) \right|^2 \ d\omega d\theta d\varphi}
	\label{eq:ASFCLR}
\end{equation}
where $Y_{m}(\omega)$ is the power spectral density of $y_m$, $E_m(\theta,\varphi)$ is the $m$th far-field component as defined in \eqref{eq:Farfield} and
\begin{equation}
	\Omega_{\text{Useful}} = \left[-\frac{\omega_{\text{BW}}}{2},\frac{\omega_{\text{BW}}}{2}\right]\times\left[\theta_l,\theta_u\right]\times\left[\varphi_l,\varphi_u\right]
\end{equation}
in which $\omega_{\text{BW}}$ is the signal bandwidth and $\theta_l,\theta_u,\varphi_l,\varphi_u$ delimit the azimuth and elevation angles of interest. $\Omega_{\text{Unwanted}}$ is the complement of $\Omega_{\text{Useful}}$.


\subsection{The Channel Model}
\label{sec:ChannelMod}

In order to analyze the link performance of a massive MU-MIMO transceiver, we need a channel model. It has been shown that user correlation in massive MU-MIMO systems can be made small if the users are sufficiently separated. 
For the purpose of studying beam-forming properties, it is crucial that the channel model contains a line-of-sight (LOS) component. 
We have therefore chosen the Rice model, which can emulate any mixture between LOS and a rich scattering environment through the $\kappa$-factor. 
According to the Rice model
\begin{equation}
	\textbf{H} = \sqrt{\frac{1}{1+\kappa}} \textbf{H}_{\text{IID}} +  \sqrt{\frac{\kappa}{1+\kappa}} \textbf{H}_{\text{LOS}}
\end{equation}
where $\textbf{H}_{\text{IID}}\sim\mathbb{C}\mathcal{N}\left(\mathbf{0},\mathbf{I}_{M\times K}\right)$ and $\textbf{H}_{\text{LOS}} = [\textbf{h}_1, \ldots, \textbf{h}_{K}]\in\mathbb{C}^{M \times K}$ with $\textbf{h}_k$ being the $K$ different LOS-vectors derived from the UE placement relative to the BS as
\begin{IEEEeqnarray}{rCL}
	\textbf{h}_k = \left[ 1, \exp\!\left(i\frac{2\pi}{\lambda} \psi_m \right), \ldots, \exp\!\left(i\frac{2\pi}{\lambda} (M-1)\psi_m\right) \right]^T\!\!. \IEEEeqnarraynumspace
\end{IEEEeqnarray}
Here, $\psi_m = \mathrm{X}_m\sin(\theta)\cos(\varphi) + \mathrm{Y}_m\sin(\theta)\sin(\varphi)$ and $\mathrm{X}_m$ and $\mathrm{Y}_m$ are the spatial coordinates of the $m$th element. Finally, $d$ is the element spacing. The vectors in the IID component are all uncorrelated. 
For the LOS component, the channel vector correlation, $\Exp{}{\textbf{h}_k\textbf{h}_l^H}$ for $k\neq l$, depends on the user placement. By increasing $\kappa$ and placing users closely together, we increase the correlation, as described in \cite[Eq. 7.35]{tse2005fundamentals}.

\subsection{The Pre-coding Scheme}

For all simulations presented in this paper, we use the regularized zero-forcing (RZF) pre-coder as defined in \cite{couillet2011random}, i.e., $\Gdl = \textbf{H}\left(\textbf{H}^H \textbf{H} + \frac{1}{\text{SNR}}\textbf{I}_M\right)^{-1}$.  We further assume that the BS transmitter has perfect knowledge of both \textbf{H} and SNR.

\section{Parametric Studies Using Deterministic Behavioral Models}
\label{sec:Paramstudy}

\subsection{Amplifier Distortion and Mutual Coupling}
\label{sec:PADist}

In this section, we analyze the impact of mutual coupling and channel correlation on the average received EVM. First, consider a memoryless\footnote{Some generality is lost when removing the memory terms.} version of the PA model in \cite{FagerPA}
\begin{IEEEeqnarray}{rCl}
	y_m[n] &=& \chi_{1}x_{m}[n] + \underbrace{\displaystyle\sum^{P_1}_{p_1=2} \chi_{p_1} x_{m}[n] |x_{m}[n]|^{2(p_1-1)}}_{d_{m,0}[n]}\nonumber\\ 
	&&+ \underbrace{\sum^{P_2}_{p_2=1} \eta_{p_2} x_{m,r}[n] |x_{m}[n]|^{2(p_2-1)}}_{d_{m,1}[n]} \nonumber\\
	&&+ \underbrace{\sum^{P_2}_{p_2=2} \gamma_{p_2} \bar{x}_{m,r}[n] x_{m}^2[n] |x_{m}[n]|^{2(p_2-2)}}_{d_{m,2}[n]} \label{eq:pamodfirst}
\end{IEEEeqnarray}
where $P_1$ and $P_2$ are the non-linear orders and $\chi_{p_1}$, $\eta_{p_2}$ and $\gamma_{p_2}$ are the model parameters.\footnote{The parameters used in the following simulations are based on measurements performed in \cite{FagerPA}.} Let $d_m[n] = d_{m,0}[n] + d_{m,1}[n] + d_{m,2}[n]$. The received signal at the $k$th user may now be calculated as\footnote{For simplicity of notation, we have left out the time index $n$ here.} 
\begin{IEEEeqnarray}{rCl}
	r_k &=& \displaystyle\frac{1}{M}\sum^{M}_{m=1} h_{k,m} \ y_m + n_k\nonumber\\
	&=& \frac{1}{M}\sum^{M}_{m=1} \left(\chi_{1} h_{k,m} \ x_m + h_{k,m} \ d_m\right)  + n_k  \nonumber\\
	&=& \chi_{1} s_k + \underbrace{\frac{1}{M}\sum^{M}_{m=1}  h_{k,m} \ d_m}_{d_r}  + n_k \nonumber\\
	&=& \chi_{1} s_k + n_k. \label{eq:pamodfin}
\end{IEEEeqnarray}
The last equality holds provided that the term $d_r$ tends to zero as $M\rightarrow\infty$. This happens when the correlation between $d_m$ and $h_{k,m}$ vanishes when $m$ grows large. It is further worth noticing that the coefficient $\chi_{1}$ of the leading term in the RF PA model does not depend on $M$.

We will now investigate the dependency of the overall distortion on the channel correlation by using the Rice model as described in Section \chapref{ChannelMod}. Specifically, we study how the average user received EVM varies as a function of increased coupled power and channel correlation. The power transfered from neighboring antennas through mutual coupling is determined by the array element spacing, which ranges from $0.35 \lambda$ to $1.5 \lambda$, and the channel correlation may be increased using the $\kappa$-factor in the Rice model.

\begin{figure}
\begin{center}
\includegraphics[width=0.4\textwidth]{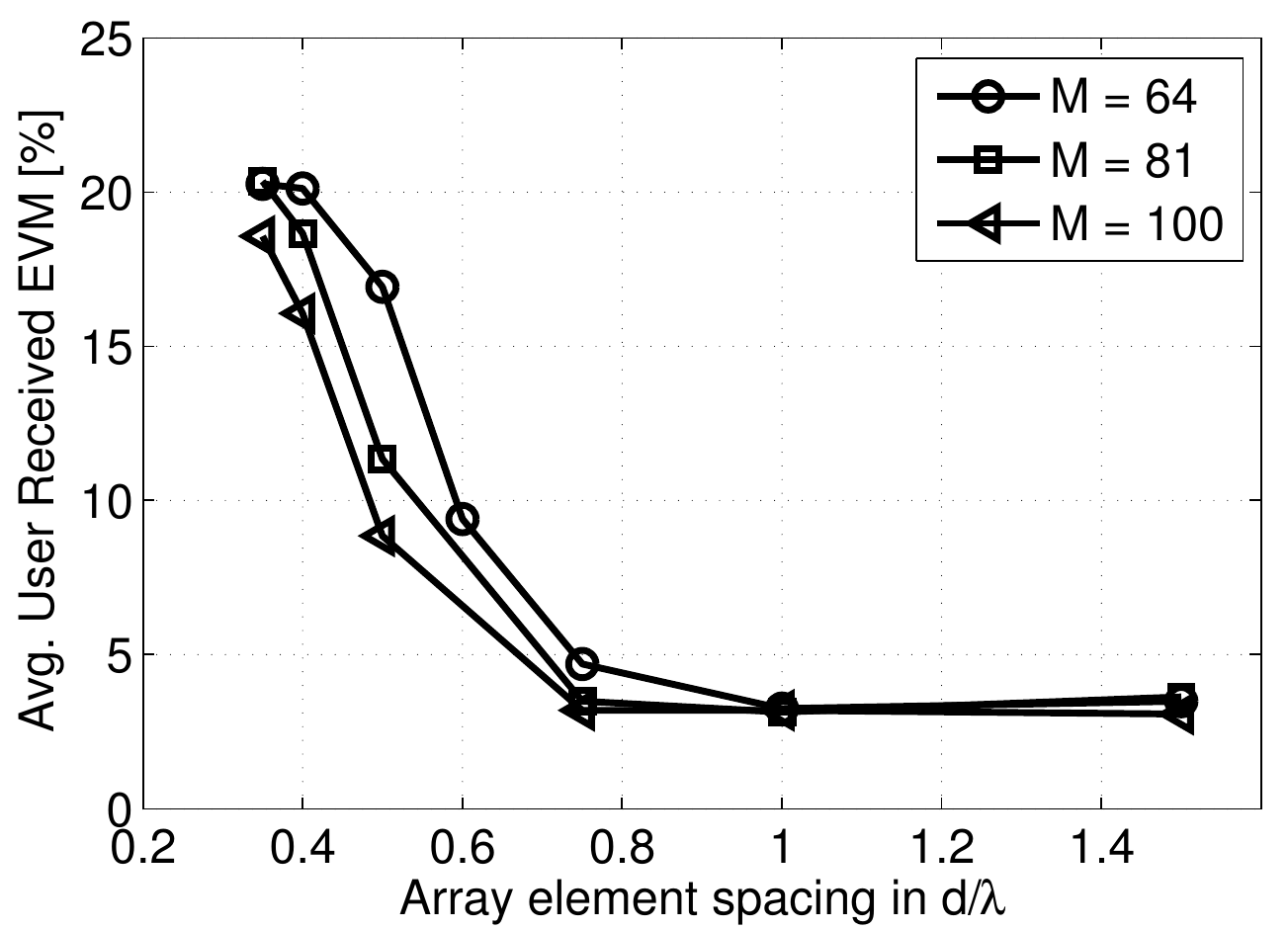}
\end{center}
\caption{Average received EVM for three different array sizes over a  LOS channel with $\kappa = 100$. This simulation contains only PA and array models.}
\label{fig:AvgEVMNr1}
\end{figure}

In order to illuminate the impact of correlation over the array, we consider the case $K = 4$, but with more closely spaced users than defined in \tabref{SimParams}. Specifically, the azimuth spacing between two of the users is fixed to $3$~degrees. \figref{AvgEVMNr1} shows the case of a channel with strong LOS component ($\kappa = 100$). We see that increasing the number of elements in the array has a minor impact on the average received EVM at large array element spacings. However, the impact of the coupling through the polynomial cross-terms, as defined in \eqref{eq:pamodfirst}, drastically increases as we increase the amount of coupling by decreasing the element spacing. 

\begin{figure}
\begin{center}
\includegraphics[width=0.42\textwidth]{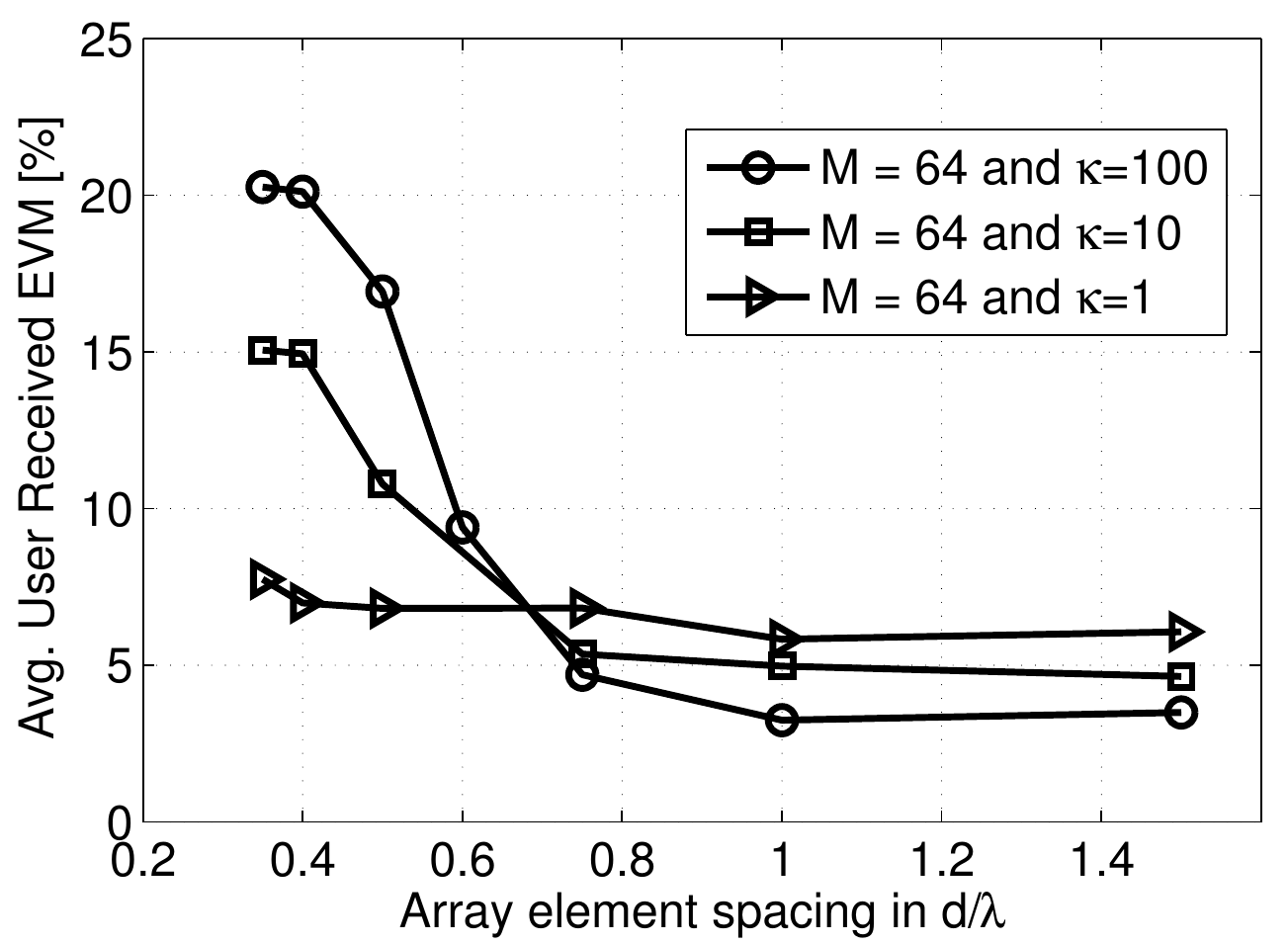}
\end{center}
\caption{Average received EVM for a 64 element array for three different mixtures of IID/LOS with $\kappa = 1, 10,$ and $100$.  This simulation contains only PA and array models.}
\label{fig:AvgEVMNr2}
\end{figure}
\begin{figure}
\begin{center}
\includegraphics[width=0.42\textwidth]{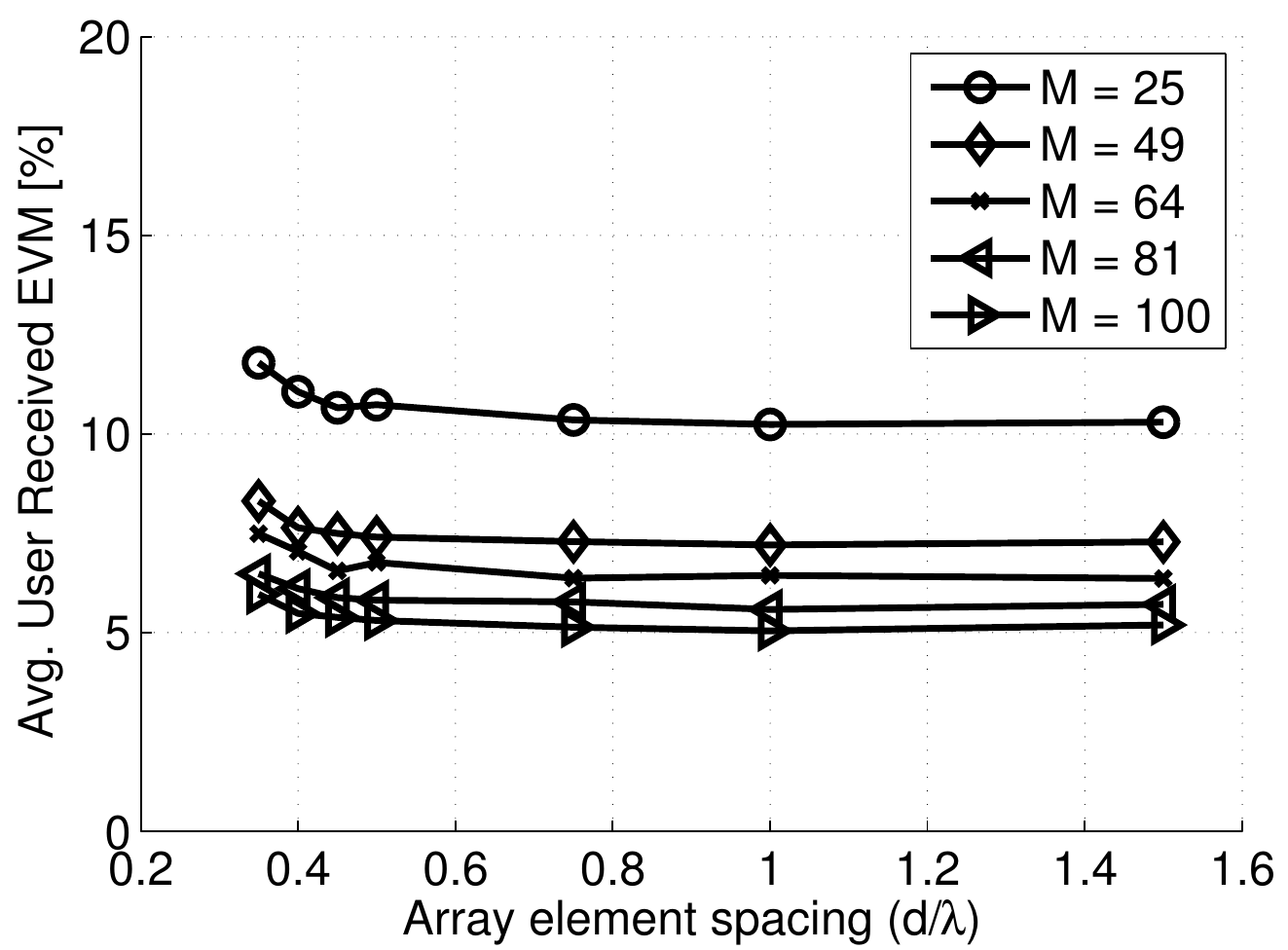}
\end{center}
\caption{Average received EVM for different array sizes over a Rayleigh channel ($\kappa = 0$).  This simulation contains only PA and array models.}
\label{fig:AvgEVMNr3}
\end{figure}

Reducing the $\kappa$-factor from $100$ to $1$, as illustrated in \figref{AvgEVMNr2}, we observe a decreased sensitivity for increased mutual coupling over the array, as predicted by \eqref{eq:pamodfirst}--\eqref{eq:pamodfin}, because the channel correlation now decreases. As we reach a pure Rayleigh channel, i.e., $\kappa = 0$, the sensitivity is so low that for a sufficiently large number of antenna elements, the average received EVM appears invariant to the mutual coupling, see \figref{AvgEVMNr3}.

\subsection{DAC Resolution}
\label{sec:DCConv}

As previously mentioned, the power consumption per DAC scales with both sample-rate and resolution. However, leveraging on the large number of degrees of freedom available in massive MIMO, one may consider the use of DACs with much lower resolution than in current wireless systems, and rely on the massive number of antenna elements to overcome the loss in resolution.

\begin{figure}
\begin{center}
\includegraphics[width=0.42\textwidth]{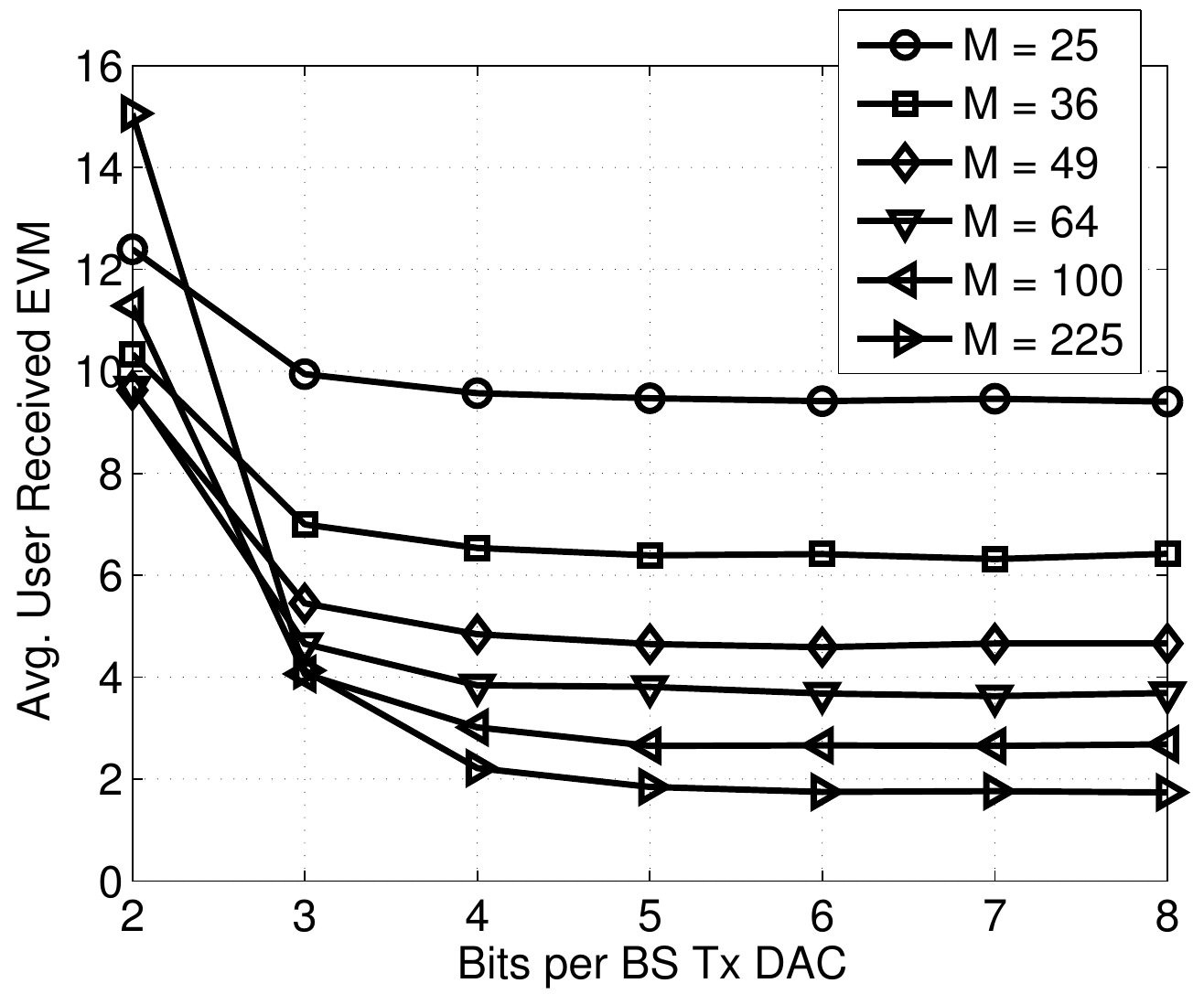}
\end{center}
\caption{Average received EVM as a function of $M$ and the number of bits per transmitter DAC. This simulation contains only DAC models.}
\label{fig:DACandArraySizeSweep}
\end{figure}

In this section, we study the impact of DAC resolution on both the average user received EVM and the transmitted unwanted emission. We consider $2$--$8$ bits per DAC and transmit chain. \figref{DACandArraySizeSweep} shows the average received EVM as a function of the number of bits per transmitter DAC for different array-sizes in the range $M\in [25,225]$.  The plot suggests the existence of a threshold after which adding more resolution does not lower the average user received EVM. This threshold is expected to depend heavily on the scenario considered (LOS/NLOS, SNR, etc\dots).

\subsection{A Novel Approach to Dithering in MIMO Systems}

\begin{figure}
\begin{center}
\includegraphics[width=0.42\textwidth]{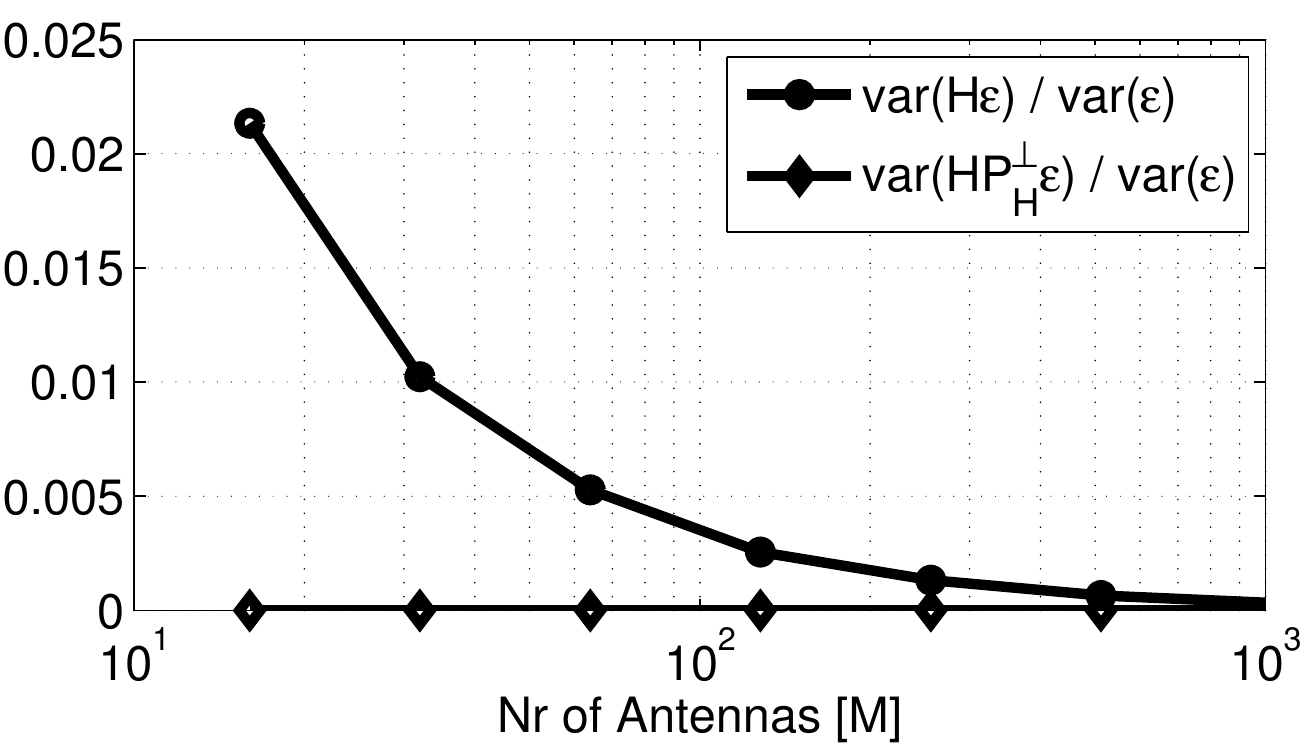}
\end{center}
\caption{The convergence rate of $\textbf{H}\boldsymbol\epsilon$ and $\textbf{H}\Pnull\boldsymbol\epsilon$ with $M$ over an IID Rayleigh channel and $K = 10$. Each point is averaged over 200 independent channel realizations.}
\label{fig:DitheringConvergenceRate}
\end{figure}

Dithering is a common tool in quantization, used to increase the effective number of bits, while keeping the converter resolution low. In multi-antenna systems, the main task of the dither is to de-correlate the quantization noise across antennas so that it is averaged out at each receiver. The most common method is to add a random vector $\boldsymbol\epsilon$ with entries $\epsilon_m\sim\mathcal{U}(-\text{LSB}/2,\text{LSB}/2)$ where LSB stands for least significant bit. As shown in \figref{DitheringConvergenceRate} for the case of the IID Rayleigh channel, in massive MU-MIMO the random dithers $\epsilon_m\sim\mathcal{U}(-\text{LSB}/2,\text{LSB}/2)$, which are independent over the antennas, cancel out as $M$ grows large. Specifically, we have that $\textbf{H}\boldsymbol\epsilon\rightarrow \textbf{0}$ as $M\rightarrow\infty$.
 
\begin{figure}
\begin{center}
\includegraphics[width=0.42\textwidth]{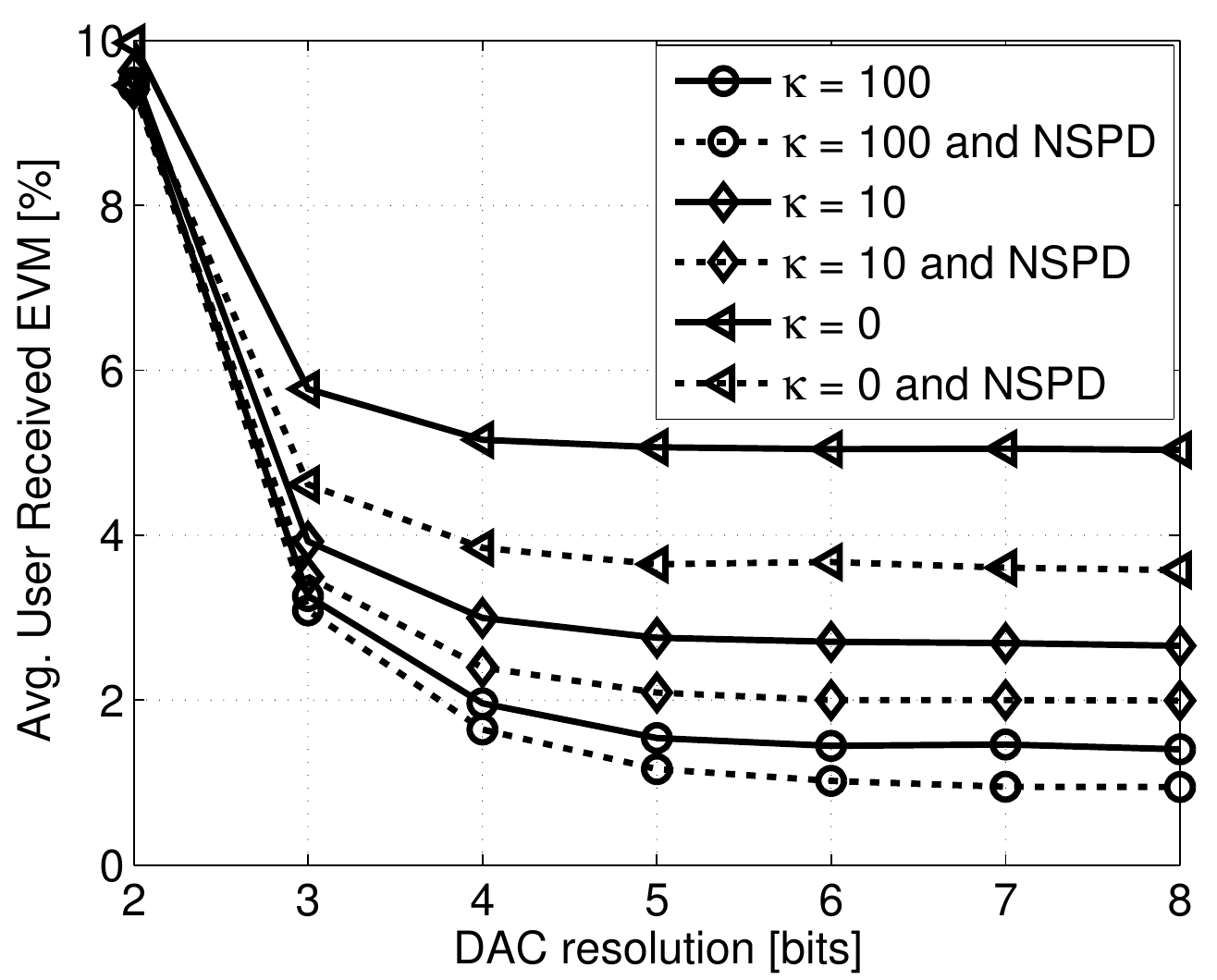}
\end{center}
\caption{Average received EVM over different mixtures of LOS/IID, without dithering and NSPD. This simulation contains only DAC models.}
\label{fig:DACDithering}
\end{figure}

In order to enforce $\textbf{H}\boldsymbol\epsilon = \textbf{0}$ even for small array sizes, we can project the dithering vector onto the channel null-space using the orthogonal projection matrix $\Pnull = \textbf{I} - \textbf{H}^H(\textbf{H}^H\textbf{H})^{-1} \textbf{H}$. This gives us
\begin{equation}
	\textbf{y} = \textbf{H}(\textbf{x} + \Pnull\boldsymbol\epsilon) = \textbf{H}\textbf{x} + \textbf{H}\Pnull\boldsymbol\epsilon = \textbf{H}\textbf{x}.
\end{equation}
Assuming perfect channel knowledge at the BS transmitter, this guarantees that $\textbf{H}\boldsymbol\epsilon = \textbf{0}$ for any $M$ as illustrated in \figref{DitheringConvergenceRate}. Simulations illustrating the impact of DAC resolution on the average received EVM are shown in \figref{DACDithering}, alongside  with the null-space projected dithering (NSPD) results. 


\subsection{Unwanted Emissions}

\begin{figure}
\begin{center}
\includegraphics[width=0.42\textwidth]{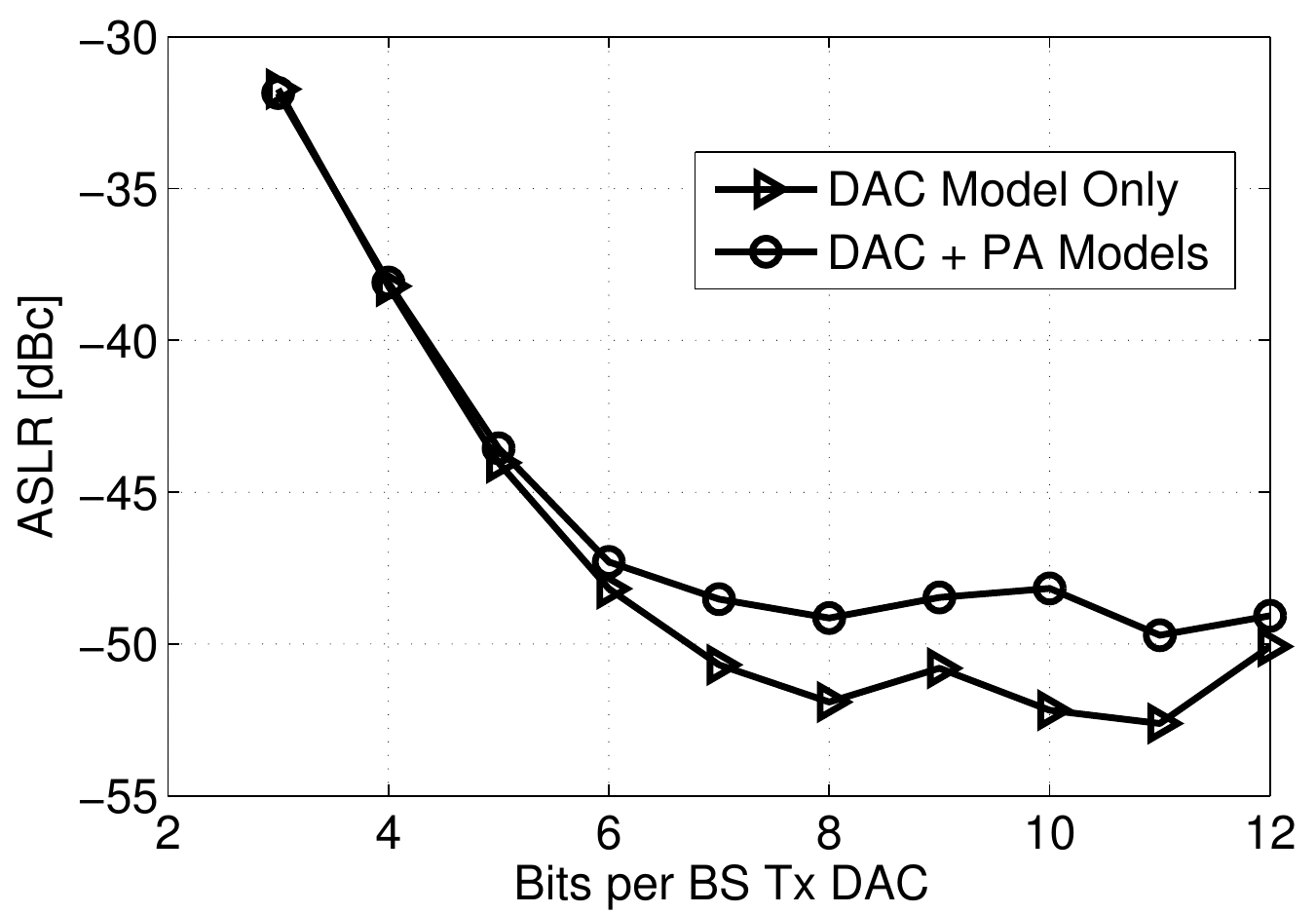}
\end{center}
\caption{Unwanted emissions versus per-antenna transmitter DAC resolution for an ideal BS with only DAC-models, and a BS with both DAC and PA/array models.}
\label{fig:OOBOODOA}
\end{figure}

We simulate the total transmitted unwanted emission power relative to the average received power per user. The results in \figref{OOBOODOA} are obtained using both an ideal BS comprising only DAC quantization noise, and a less ideal BS, which further comprises PAs and array mutual coupling effects. Also in this case, we can observe the existence of a resolution threshold.

\section{Comparing Modeling Approaches - Stochastic Versus Deterministic Models}
\label{sec:Combined}

So far we have studied each impairment as a separate phenomenon. Now we study their combined effects and compare the results to ones predicted by the simplified stochastic models presented in \eqref{eq:AddMod} and \eqref{eq:MultMod}.
Throughout this section, we set the array spacing to $d = 0.5\lambda$. The parameters of the stochastic components in each model ($\nu$ for the additive model and $\sigma^2_a$ and $\sigma^2_{\phi}$ for the multiplicative model) are numerically chosen to fit the results obtained with the deterministic behavioral model for the case $M = 4$.

\begin{figure}
         \centering
         \begin{subfigure}{0.42\textwidth}
                 \centering
                 \includegraphics[width=\textwidth]{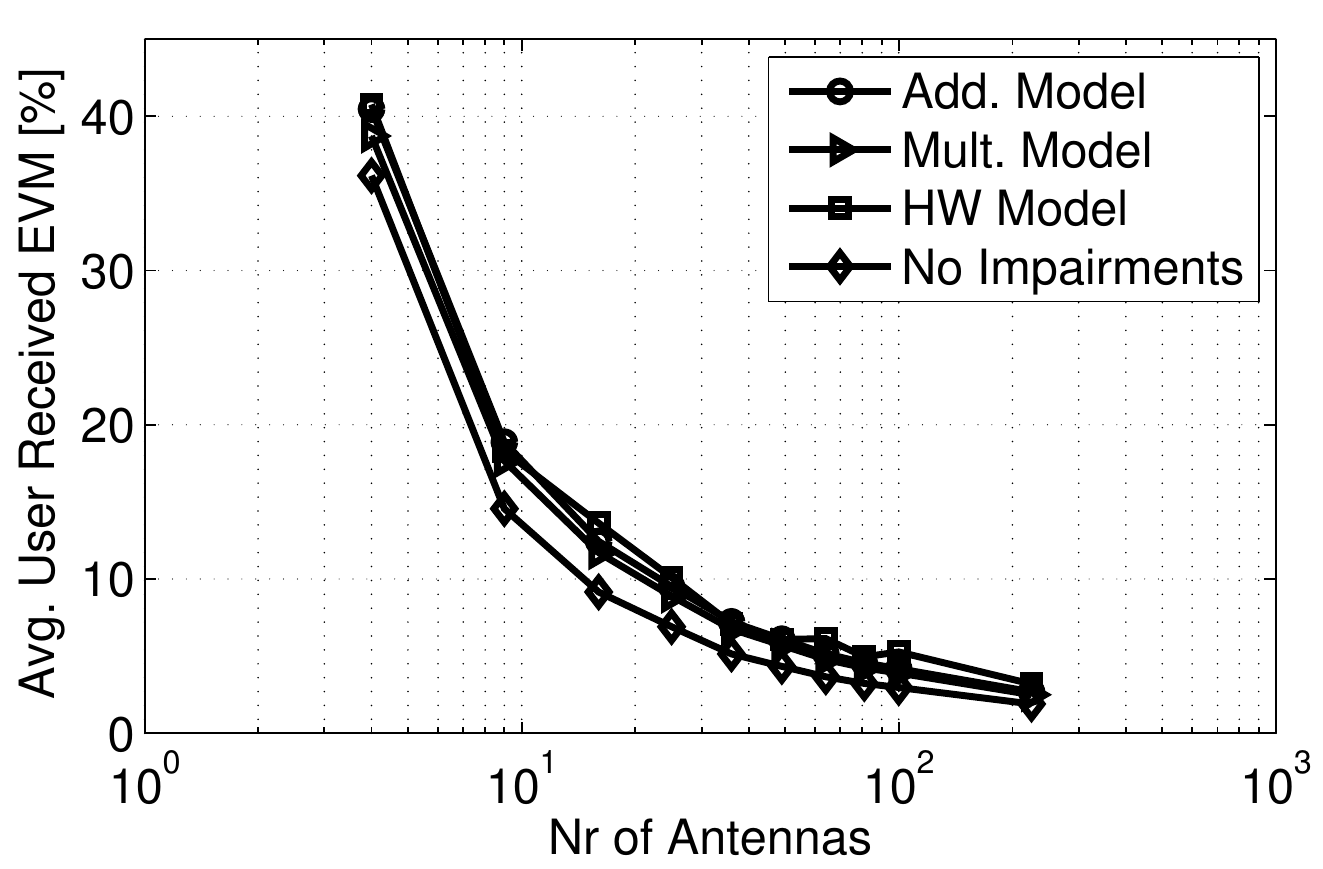}
                 \caption{ }
                 \label{fig:combined}
         \end{subfigure}%

         \begin{subfigure}{0.42\textwidth}
                 \centering
                 \includegraphics[width=\textwidth]{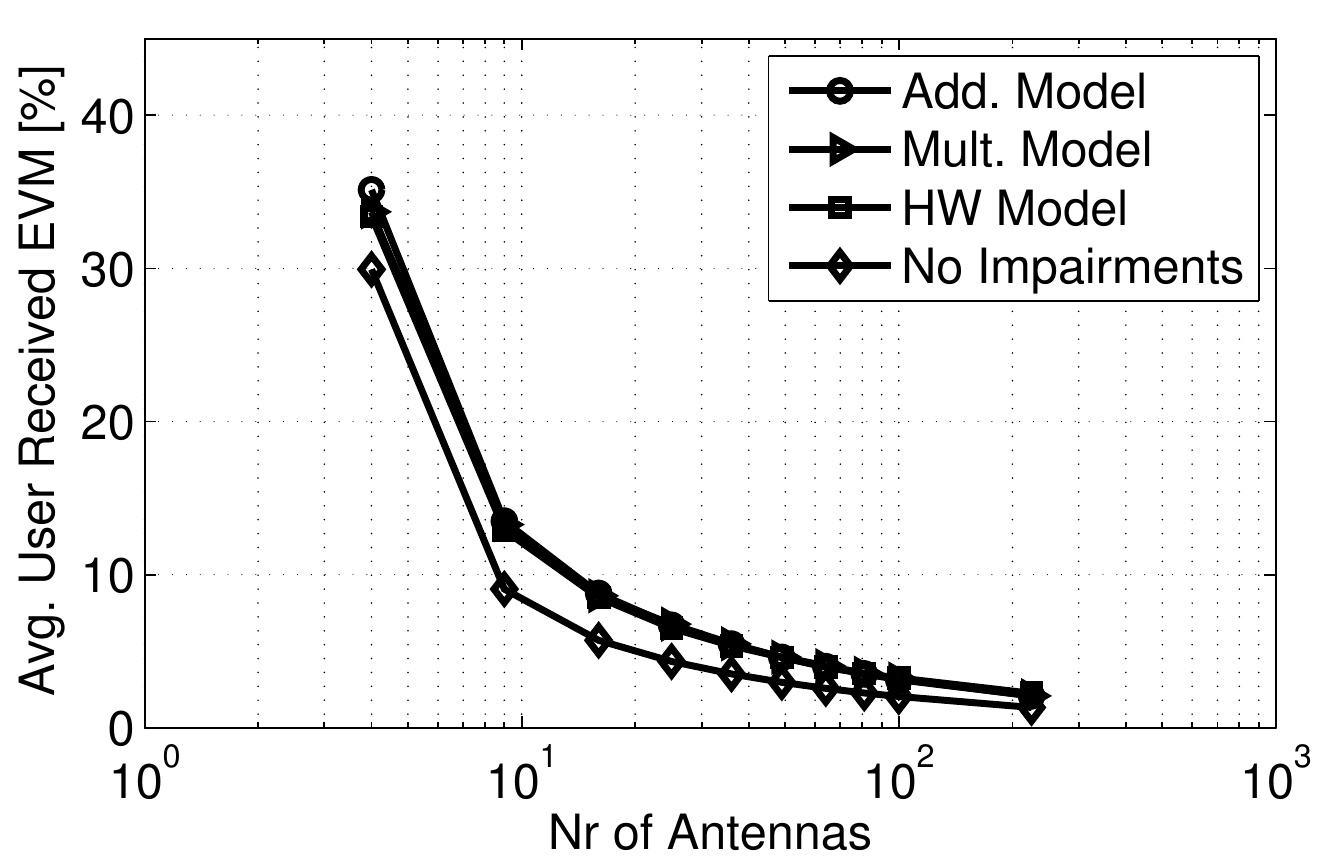}
                 \caption{ }
                 \label{fig:combinedK100}
         \end{subfigure}
         \caption{A comparison of both the statistical and deterministic models, including PA, array and 6-bit DAC models, over (a) an IID Rayleigh channel ($\kappa = 0$) and (b) a LOS channel ($\kappa = 100$). Each point is averaged over 200 channel realizations.}
         \label{fig:CombEff}
\end{figure}

The simulations shown in \figref{CombEff} suggest a good fit for both the additive and multiplicative statistical models under different channel conditions. Indeed, they yield very similar scaling behavior of the EVM as a function of the array size.

\subsection{A Note On Stochastic Modeling and Power Normalization}

As demonstrated in the previous section, the stochastic approach to modeling HW impairments and their impacts on massive MIMO appears to be sufficiently accurate. It is, however, crucial that the power normalization factors $\alpha$ and $\beta$ are included in the model. If left out, the impairments would add power, causing the model to behave nonphysically.

\begin{figure}
\begin{center}
\includegraphics[width=0.42\textwidth]{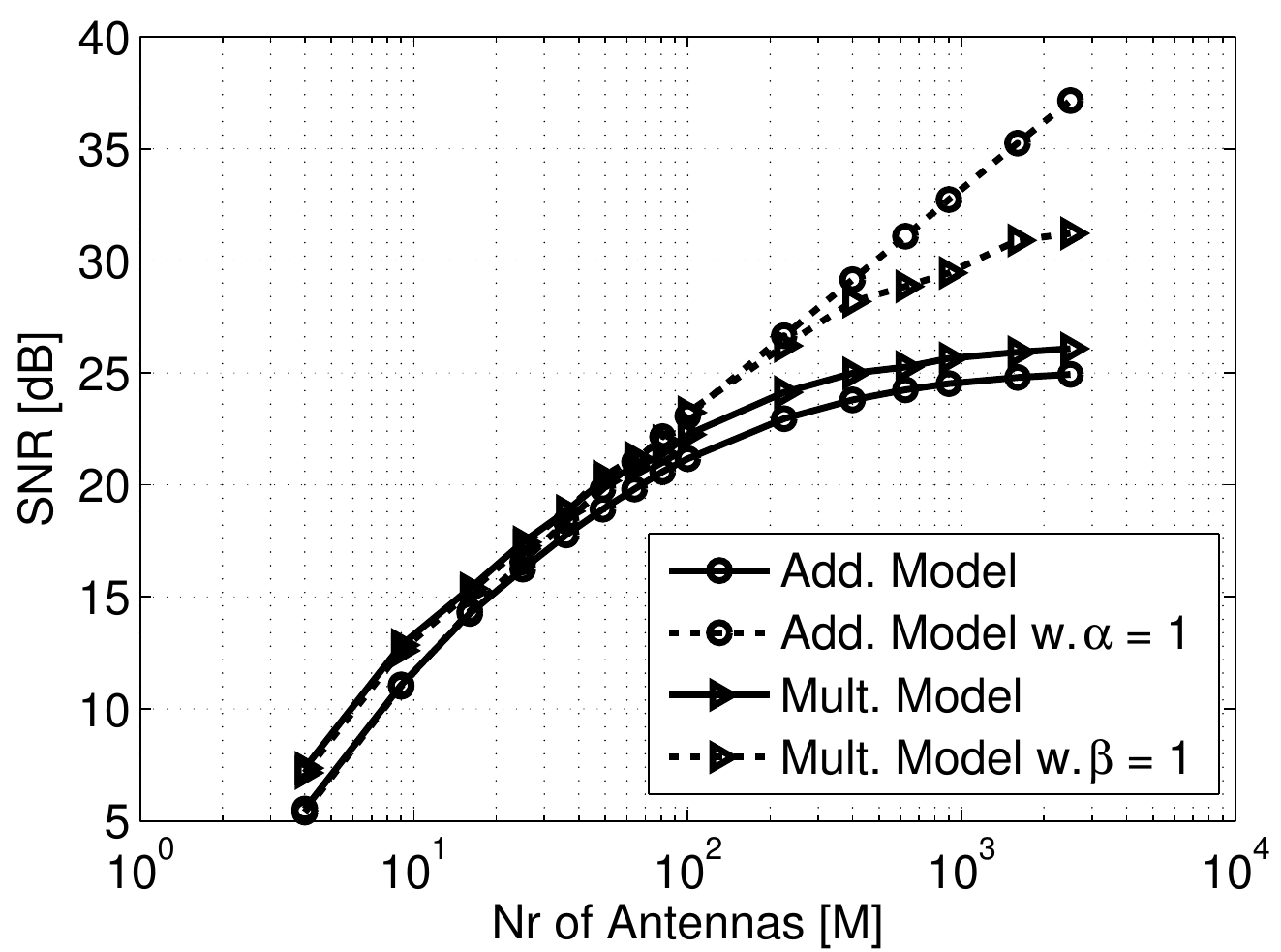}
\end{center}
\caption{SNR versus $M$ for the additive and multiplicative stochastic models, with and without power normalization.}
\label{fig:StochModSNR}
\end{figure}

It is interesting to note that the  EVM in \eqref{eq:EVMDef}  is insensitive to whether this power normalization is performed or not. However, the same does not hold for the SNR defined as
\begin{equation}
	\text{SNR} = 10\log_{10}\!\left(\frac{\Exp{}{\|\textbf{s}\|_2^2}}{\Exp{}{\|\textbf{s}-\ydl\|_2^2}}\right).
	\label{eq:SNR}
\end{equation}
As illustrated in \figref{StochModSNR}, the SNR scales linearly with $M$ for the case of the additive model without power normalization. When the power normalization is performed, the SNR converges instead to a finite value (which depends on the level of the impairments) for both the additive and the multiplicative stochastic models. This behavior can be easily explained in terms of lost power---a quantity that does not vanish with $M$, as predicted by \eqref{eq:pamodfin}.

\section{Conclusion}
\label{sec:conclusions}

We examined the impact of hardware impairments on the massive MU-MIMO performance in terms of average user received EVM and unwanted space-frequency emissions for different channel conditions, by means of both statistical as well as deterministic hardware models. Our simulation results characterize the impact of mutual coupling on the overall power amplifier distortion and its dependency on the channel correlation. These results point toward a decreased sensitivity per PA over a Rayleigh channel due to the low correlation between the coupled and the transmitted signals.

We demonstrated that low-resolution DACs decrease the overall power consumption while still maintaining low average user received EVM and controlled unwanted emissions. Moreover, we proposed a novel dithering design method that exploits the vast channel null-space available in massive MIMO systems in order to improve the link quality per user, while decreasing the number of bits per transmitter DAC.

Our results suggest that state-of-the-art stochastic models are, despite their simplicity, useful tools for the analysis of massive MU-MIMO systems. Finally, the results shown in this paper confirm that in massive MU-MIMO systems there is room for relaxing the otherwise stringent hardware requirements commonly in place on current MIMO systems.

\vfil\eject
\bibliographystyle{IEEEtran}


\end{document}